\pdfoutput=1
\documentclass[aps,prl,twocolumn,superscriptaddress,showpacs,showkeys]{revtex4-1}

\usepackage{graphicx}
\usepackage{textcomp}
\newcommand{\td}{\text{d}}
\newcommand{\st}[1]{_{\text{#1}}}
\usepackage{amsmath}

\begin{document}


\title{Attosecond pulse shaping using partial phase matching}


\author{Dane R. Austin}
\email[]{d.austin@imperial.ac.uk}
\altaffiliation{Current address: Imperial College London, London SW7 2AZ, UK}
\affiliation{ICFO-Institut de Ciences Fotoniques,  08860 Castelldefels (Barcelona), Spain}
\author{Jens Biegert}
\affiliation{ICFO-Institut de Ciences Fotoniques,  08860 Castelldefels (Barcelona), Spain}
\affiliation{ICREA-Instituci\'{o} Catalana de Recerca i Estudis Avan\c{c}ats, 08010 Barcelona, Spain}


\date{\today}

\begin{abstract}
We propose a method for programmable shaping of the amplitude and phase of the XUV and x-ray attosecond pulses produced by high-order harmonic generation. It overcomes the bandwidth limitations of existing spectral filters and enables removal of the intrinsic attosecond chirp as well as the synthesis of pulse sequences. It is based on partial phase matching, such as quasi-phase matching, using a longitudinally addressable modulation. 
\end{abstract}




\maketitle

Coherent extreme ultraviolet (XUV) and soft x-ray radiation, produced by high-order harmonic generation (HHG) in an intense laser field, is central to attoscience \cite{Nisoli-2009-New,*Krausz-2009-Attosecond}. Spectra spanning 200-1600\,eV, with the potential to support temporal features of 2.5\,as duration, have been generated, providing the raw material for coherent excitation of atomic scale electron dynamics down to the inner shell. 
At present, there is no general means of controlling the spectral and temporal profile of radiation produced via HHG on the attosecond timescale. Bandpass and dispersive filtering, the latter being necessary to eliminate the attosecond chirp that is intrinsic to HHG, have been demonstrated using metal films, gases, and multilayer mirrors, but these lack tunability and have been demonstrated below 150\,eV for bandwidths of $\approx 50$\,eV \cite{Goulielmakis-2008-Single-Cycle,*Hofstetter-2011-Attosecond,*Ko-2010-Attosecond}. 

The ability to arbitrarily shape HHG over its entire bandwidth would improve existing experiments and enable others, analogous to the role of dispersion control \cite{Walmsley-2001-role} and pulse shaping \cite{Weiner-2000-Femtosecond} in femtosecond science and technology. Using harmonic spectra from standard IR sources, pulses as short at 2.5\,as could be generated \cite{Popmintchev-2012-Bright}. Current XUV-pump IR-probe studies of single-photon ionization \cite{Schultze-2010-Delay,*Cavalieri-2007-Attosecond} could be performed over a range of photon energies, accessing a wider range of initial states and potentially disentangling the roles of the Coulomb potential and the IR probe. Coherently controlled wavepackets \cite{Remetter-2006-Attosecond} could be launched and probed. With sufficient intensity, XUV-pump XUV-probe \cite{Hu-2006-Attosecond} spectroscopy and coherent control could be achieved.

Macroscopic effects --- the coherent sum of the dipole response of all the target atoms --- play a crucial role in attosecond pulse generation via HHG \cite{Gaarde-2008-Macroscopic}. Of primary importance is the wave vector mismatch $\Delta k$ between the laser-driven dipole excitation and the propagating harmonics. Whilst the latter propagate very close to $c$, the former is influenced by diffraction in the focus or waveguide, dispersion of the neutral gas and free-electron plasma, and the intensity dependence of the electron in the continuum. Partial phase matching \cite{OKeeffe-2012-Quasi,*Zhang-2007-Quasi-phase-matching}, achieved with a longitudinal modulation in the dipole excitation of wavevector $K = \Delta k$, can be used to overcome a phase mismatch. Partial phase matching is inherently dependent on the harmonic frequency and hence offers a degree of control over the spectrum. Here, we show that using partial phase matching with a longitudinally addressable modulation, one may compensate the attosecond chirp and also synthesize an arbitrary in-situ amplitude and phase filter  for HHG. 
The method is applicable over the whole spectrum up to kiloelectron-volt photon energies. 

The essence of our method is that in HHG, a phase velocity mismatch is almost always accompanied by a group velocity mismatch i.e. $\Delta k(\omega)=\Delta n \omega / c$, where $\Delta n=n(\omega_1)-n(\omega)$ is the difference between the refractive indices at the fundamental frequency $\omega_1$ and the harmonics $\omega$.
This is because in HHG, the group delay of the constituent attosecond bursts is dictated by the laser field oscillations (rather than their envelope as with perturbative harmonic generation). The phase velocity of the laser is therefore imparted on the group velocity of the dipole response. Partial phase matching with wavenumber $K$ occurs at a single frequency 
$\omega=Kc/\Delta n$. If $K$ varies along the propagation axis, then multiple frequencies are phase matched, but because of the group-velocity mismatch, their group delays will differ, producing a relative chirp between the dipole excitation and the macroscopically generated field. A linear variation $K_1 = \td K / \td z$ leads to a quadratic spectral phase of 
$(\Delta n/c)^2/K_1$, tunable in both magnitude and sign through $K_1$. As we will show, this effect can compensate the attosecond chirp leading to transform limited pulses, or be generalized to enable arbitrary pulse shaping. 

\begin{figure}[h]
\includegraphics{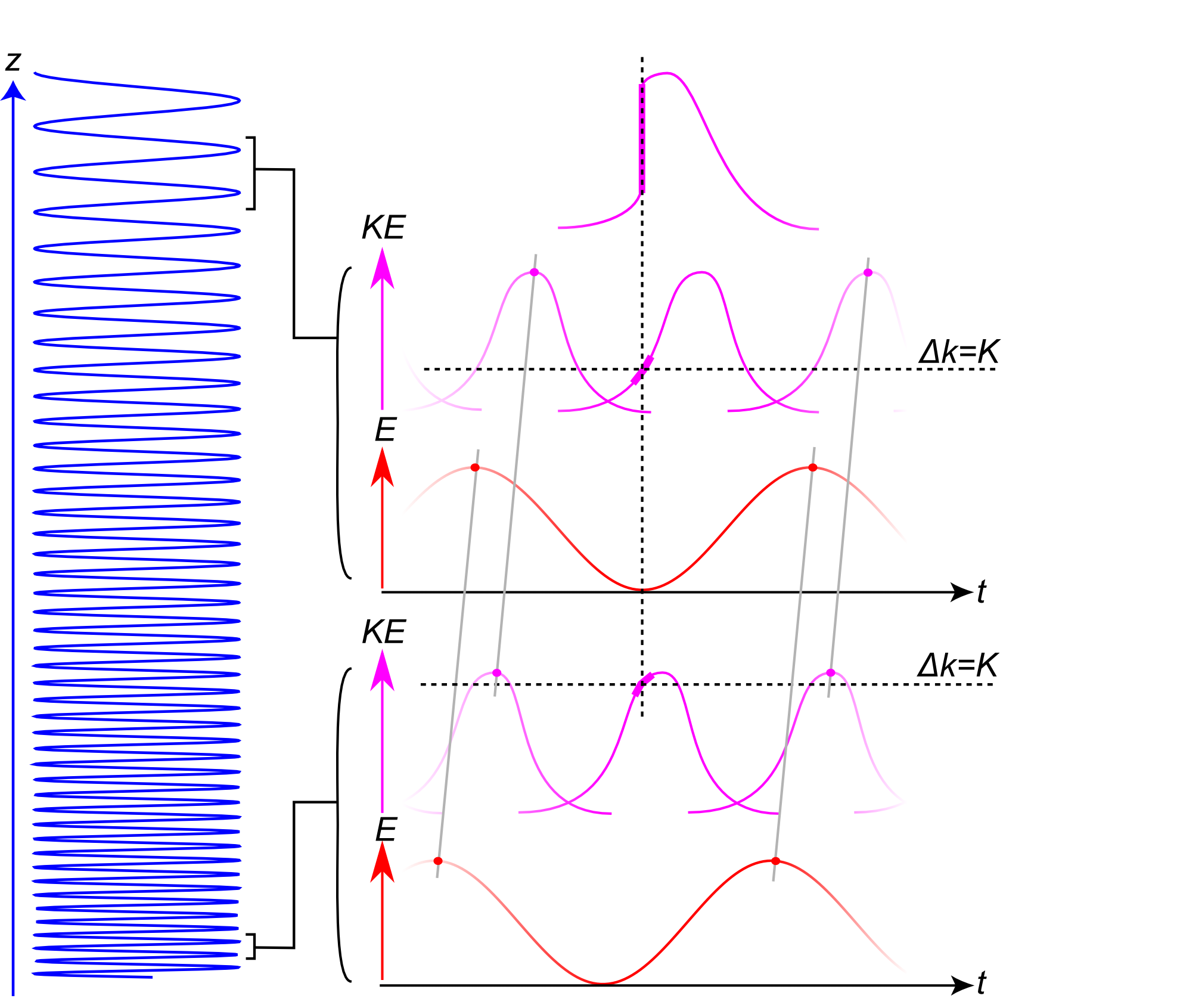}
\caption{\label{fig:diagram}Attosecond chirp removal through negatively chirped partial phase matching; longitudinal modulation (blue), laser field (red) and electron kinetic energy (purple) versus retarded time, frequency of partial phase matching (horizontal dashed lines), and recombination time of partially phase matched harmonics (vertical dashed line).}
\end{figure}

Figure~\ref{fig:diagram} is a cartoon illustration of the concept, depicting compensation of the positive chirp of the short trajectories assuming a subluminal laser phase velocity. For simplicity, we assume $n(\omega)=1$. At two points along the propagation axis $z$, the laser field $E$ (red) and the kinetic energy KE (purple) of the recombining electron are plotted versus time  in the retarded frame $t=\bar{t}-z/c$, where $\bar{t}$ is time in the lab frame. The laser field and the electron motion that it drives are delayed upon propagation, as shown by the sloped grey lines through the laser field peaks (red dots) and classical cutoffs (violet dots). A negatively chirped modulation (blue) achieves partial phase matching at decreasing harmonic frequencies, indicated by the horizontal dashed lines. The negative chirp is chosen such that the recombination time of the phase matched harmonics is constant (vertical dashed line), resulting in unchirped emission from the short trajectory.

\begin{figure}
\includegraphics{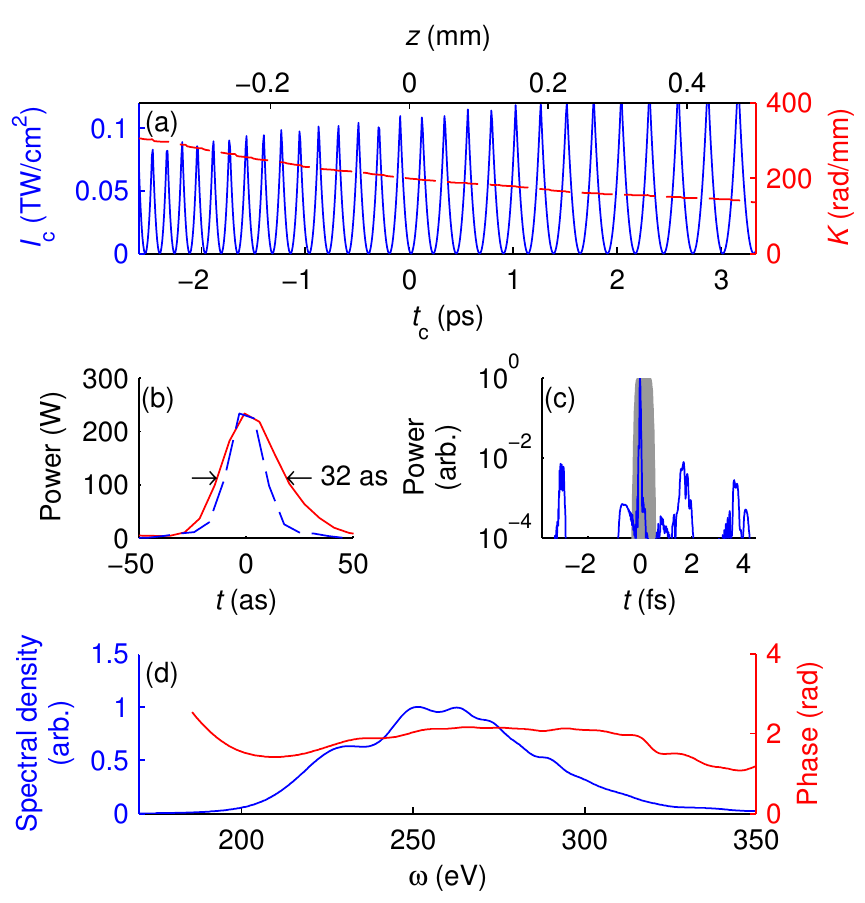}
\caption{\label{fig:TLP}Production of a transform limited 32\,as pulse with a few-cycle infrared drive field and chirped quasi-phase matching.(a) Temporal intensity of the counterpropagating pulse train (blue solid, left $y$-axis) and corresponding longitudinal spatial frequency (red dashed, right $y$-axis). (b) Instantaneous power of the generated pulse (red solid), with indicated FWHM duration and Fourier-transform limited pulse (blue dashed). (c) Same as (b) but zoomed out and on logarithmic scale. The temporal window used for computing spectra is shown in grey. (d) Spectral density (blue, left $y$-axis) and phase (red, $y$-axis).}
\end{figure}

We now present a simulation of a concrete implementation. Our laser propagation code includes dispersion, diffraction, self-phase modulation, and plasma dephasing and absorption. The single-atom response HHG code is an augmented Lewenstein model, with ADK ionization rates and photorecombination cross sections \cite{Austin-2012-Strong}. The propagation of the harmonics includes diffraction, and absorption and dispersion by the neutral gas. All quantities (including those pertaining to quasi-phase matching) have full spatio-temporal dependence with cylindrical symmetry. A temporally and spatially Gaussian driving pulse with 9\,fs full-width at half maximum (FWHM) duration, 1.8\,\textmu m center wavelength and 220\,\textmu J energy is focused to a $e^{-2}$ radius of 50\,\textmu m a distance 1.4\,mm before a jet of helium with peak pressure 5\,bar and  FWHM thickness 1.4\,mm. The retardation of the fundamental field by the neutral gas causes a positive phase mismatch. A counter-propagating pulse train (CPT) of wavelength $\lambda\st{c}=800$\,nm and total energy 15\,\textmu J 
is focused to a $e^{-2}$ radius of 50\,\textmu m in the middle of the gas jet. The temporal intensity profile $I\st{c}=|E\st{c}^2|/(2Z_0)$ of the CPT is plotted against its own comoving time $t\st{c}=\bar{t}+z/c$ in Fig.~\ref{fig:TLP}(a). The relative phase between a peak of the drive field and the CPT field varies as $\gamma=4\pi z/\lambda\st{c}$. To lowest order, the CPT perturbs the action accumulated by the electron in the continuum, resulting in a phase shift $\alpha=\alpha_0 |E\st{c}| \cos \gamma$ \cite{Cohen-2007-Optimizing}, where $\alpha_0$ is given by first-order perturbation of the SFA action integral \cite{Dudovich-2006-Measuring}. Averaging the induced phase modulation $e^{i \alpha}$ over the longitudinally rapid variation of $\gamma$, the CPT is found to modulate the harmonic emission by a factor $J_0(\alpha_0 |E\st{c}|)$ where $J_0$ is the zeroth-order Bessel function \cite{Cohen-2007-Optimizing}. The slow variations of $|E\st{c}|$ are experienced by the drive field as a longitudinal modulation of frequency $K=4\pi/(cT\st{c})$, overlaid in Fig.~\ref{fig:TLP}(a). Here, $T\st{c}$ is the (longitudinally varying) period of the CPT. The decrease of $K$ with $z$ induces quasi-phase matching at a decreasing harmonic frequency in the manner of Fig.~\ref{fig:diagram}. The generated macroscopic field is passed through a 100\,nm silver spectral filter and a 0.5\,mrad radius far-field spatial filter to eliminate the long trajectories. The resulting temporal profile is shown in Fig.~\ref{fig:TLP}(b). A 31\,as pulse is produced, quite close to the transform limited duration of 20\,as. Note that all the temporal and spectral profiles in this Letter are radially integrated to infinity (corresponding to the experimental observable in e.g. photoelectron spectroscopy), proving the absence of significant spatio-temporal distortion. Temporal gating of the emission to a single half-cycle is achieved by the short pulse duration, following Goulielmakis et al. \cite{Goulielmakis-2008-Single-Cycle}. However, because of the spectral selectivity of the quasi-phase matching, a high-pass spectral filter is not needed. To verify that an isolated pulse is produced, Fig.~\ref{fig:TLP}(c) shows a zoomed out temporal profile on a logarithmic scale. Satellite pulses are at the 1\% intensity level, below the detection threshold of current experiments, and the main pulse contains $>90$\% of the energy. The spectral density and phase, the latter being an intensity-weighted radial average, are shown in Fig.~\ref{fig:TLP}(d). A maximum phase deviation of 0.6\,rad across the full width at 10\% bandwidth 204--310\,eV shows that the attochirp has been completely compensated. Note that in Fig.~\ref{fig:TLP}(d), as with the other spectra in this Letter, a numerical window, shown in grey in Fig.~\ref{fig:TLP}(c), has been applied to the temporal profile to prevent the weak satellite pulses from causing fine interference fringes which obscure the key result.

\begin{figure}
\includegraphics{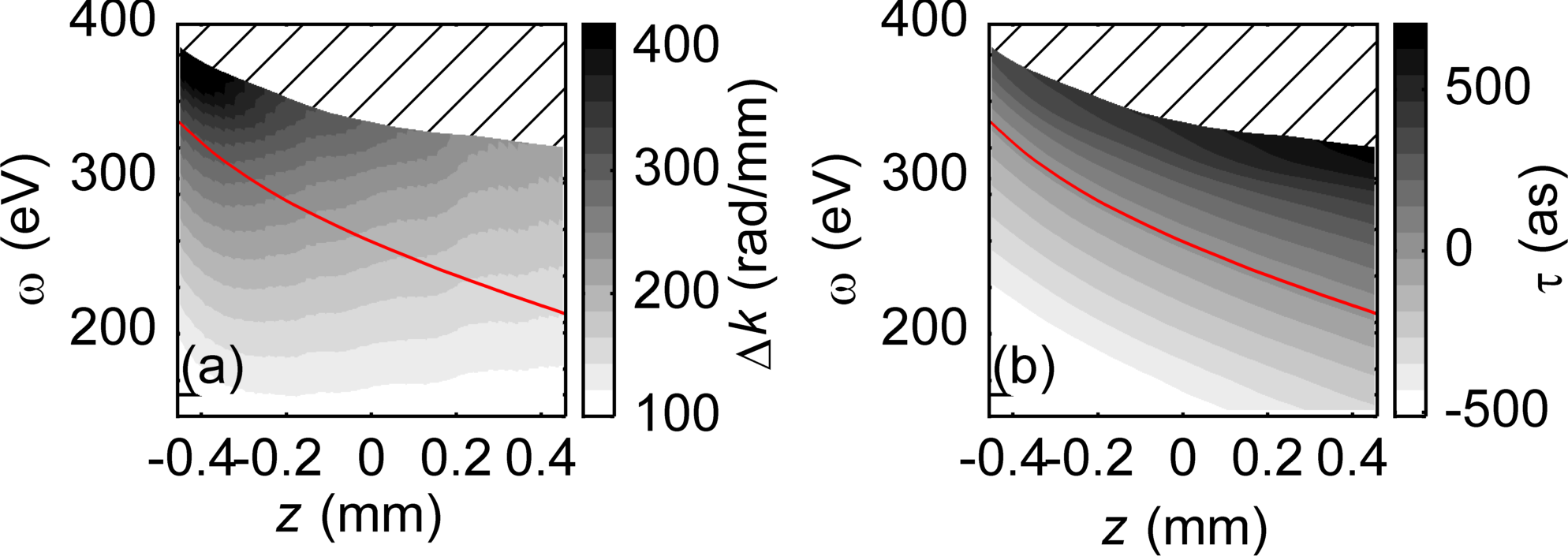}
\caption{\label{fig:TLP_Deltak}(a) Phase mismatch and (b) group delay of short trajectories versus harmonic frequency and propagation distance. The reference for the group delay is defined at  $(z,\omega)=$(0\,mm,259\,eV). The red lines show the group delay contour $\tau=0$. The hatched area is above the classical cutoff, with no meaningful phase.} 
\end{figure}

We now examine the design of the CPT, Fig.~\ref{fig:TLP}(a), in more detail. We focus on the emission from the short trajectory by the dominant laser half-cycle. Figure~\ref{fig:TLP_Deltak}(a) shows the phase mismatch $\Delta k$  extracted from the simulation in the $(\omega,z)$ domain. It shows that the phase mismatch is not perfectly proportional to the harmonic frequency, and varies with $z$. These departures from the simple picture leading to Fig.~\ref{fig:diagram} are due to the intensity-dependent dipole phase, the Gouy phase, plasma-induced dephasing of the fundamental and (weakly) the dispersion of the neutral gas at the XUV frequencies. They necessitate a slight refinement. Our method is to extract the group delay $\tau(\omega,z)$ of the harmonics, including propagation to the end of the gas. This is shown in Fig.~\ref{fig:TLP_Deltak}(b). Starting at $(\omega_0,0)$, where $\omega_0=259$\,eV is a chosen center frequency, we trace out a contour $(\omega \st{P}(z),z)$ of $\tau$. The required longitudinal spatial frequency of the CPT is the phase mismatch evaluated along this contour i.e. $K(z)=-\Delta k(\omega\st{P}(z),z)$. The red lines superimposed on Fig.~\ref{fig:TLP_Deltak}(a) and (b) illustrate this process. The amplitude $|E\st{c}(t\st{c})|$ of the CPT is then
\begin{equation}
\left| E\st{c}\left(\frac{2z}{c}\right)\right|=f[\Phi(z),\epsilon]/\alpha_0(\omega\st{P}(z),z) \label{eq:Ec}
\end{equation}
where $\Phi(z)=\int K(z) \td z$ is the phase of the CPT, $\alpha_0(\omega\st{P},z)$ the frequency-dependent phase perturbation coefficient defined above, and $f(\Phi,\epsilon)$ is chosen to map the full range of the zeroth order Bessel function onto a sinusoid of phase $\Phi$ and amplitude $\epsilon$:
\begin{equation}
f(\Phi,\epsilon)=J_0^{-1} \left[ \epsilon (0.701\cos \Phi + 0.299) \right].
\end{equation}
The parameter $\epsilon$ is used below to extend the method to amplitude modulation; for now, we take $\epsilon=1$. The particular form (\ref{eq:Ec}) gives optimal QPM efficiency and minimizes unwanted spatial overtones, but is not essential. In an experiment, $\tau(\omega,z)$ and $\Delta k(\omega,z)$ could be obtained in a calibration step, with trial CPT sequences generated by a programmable femtosecond pulse shaper \cite{Weiner-2000-Femtosecond}.

\begin{figure}
\includegraphics{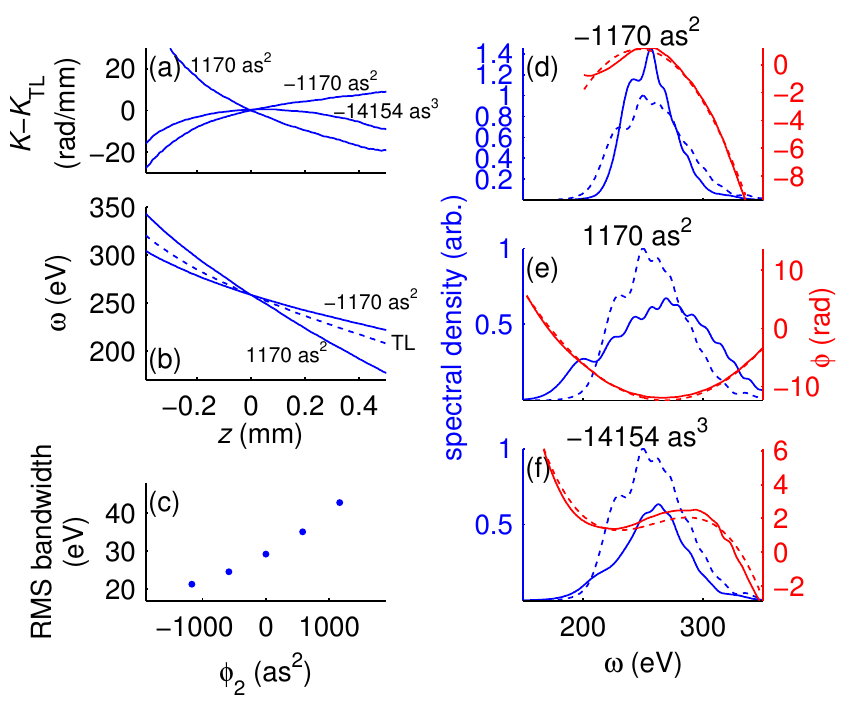}
\caption{\label{fig:chirp}(a) QPM spatial frequency relative to transform-limited case for generation of labelled spectral phases. (b) Phase-matched harmonic frequency for generation of the labelled quadratic spectral phases and the transform limited case (dashed). (c) Root-mean square bandwidth of the generated pulse versus generated quadratic spectral phase coefficient. (d)-(f) Spectral density (blue, left $y$-axis) and phase (red, right $y$-axis) of the resulting attosecond pulses corresponding to (a).  The spectral density for the transform-limited case is shown for reference by the dotted blue lines. The spectral phases are blanked out at the 1\% intensity level. The target spectral phases, with zeroth- and first-order terms adjusted for best fit, are shown by dashed red lines.}
\end{figure}

The preceding method may be generalized to produce an arbitrary smoothly varying group delay $\tau\st{S}(\omega)$ by tracing a contour of $\Delta \tau = \tau(\omega,z)-\tau\st{S}(\omega)$. All the subsequent steps are identical. Figure~\ref{fig:chirp}(a) shows the CPT spatial frequencies, relative to the transform limited case, required to produce a quadratic spectral phase of $\pm1170\,$as$^2$ ($\pm 2$\,atomic units) and a cubic spectral phase of -14154\,as$^3$ (-1 atomic unit). Qualitatively, the curves correspond to the imparted group delay: approximately linear (around $z=0$) for the quadratic spectral phase, and quadratic for the cubic spectral phase. The phases of the resulting attosecond pulses, along with the target phases are shown in Fig.~\ref{fig:chirp}(d),(e) and (f). There is excellent agreement between the actual and target spectral phases. However, the bandwidth of the chirped pulses is different to the transform limited case, shown by the dashed blue lines. This coupling of chirp to bandwidth is caused by clipping of the $z$-dependent phase matched frequency and is inherent to the method. Figure~\ref{fig:chirp}(b) compares the phase-matched frequency versus $z$ of the chirped and transform limited cases. For a positively chirped pulse, closer to the intrinsic attosecond chirp of the dipole response, the phase-matched frequency sweep is faster, and a larger bandwidth is generated within the interaction region. The opposite applies for a negatively chirped pulse. The trend is illustrated in Fig.~\ref{fig:chirp}(c), which shows the root-mean square bandwidth versus quadratic spectral phase. In general, the interaction length cannot be arbitrarily increased; plasma-induced defocusing reduces the intensity of the fundamental and the neutral gas absorbs harmonics generated at the start. This sets the ultimate limit to the pulse shaping capability of the method.

\begin{figure}
\includegraphics{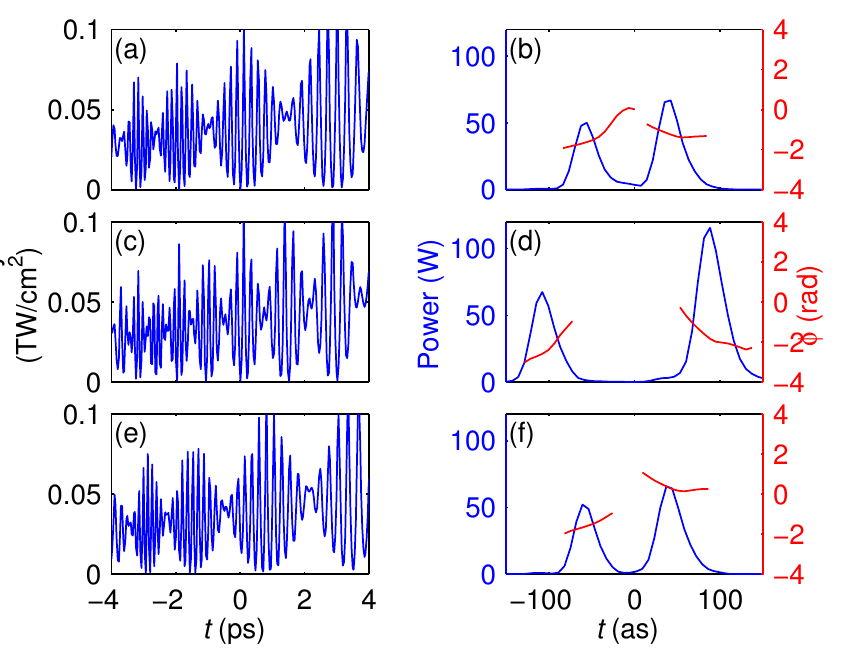}
\caption{\label{fig:delay} (a),(c),(e) Counterpropagating pulse trains for the generation of isolated double attosecond pulses (b),(d),(f). (a),(b): 97 as separation. (c),(d):  194 as separation. (e),(f) 97 as separation with $\pi/2$ phase shift on the second pulse.}
\end{figure}

A further generalization is to exploit the one-to-one correspondence between $z$ and $\omega$ by directly modulating the phase and amplitude of the CPT, producing an arbitrary transfer function $H(\omega)$. In (\ref{eq:Ec}), one sets $\Phi(z) \rightarrow \Phi(z) - \arg H(\omega\st{P}(z))$ and $\epsilon=|H(\omega)|$. In this sense, the method is analogous to programmable acousto-optic filters used for femtosecond pulse shaping \cite{Verluise-2000-Amplitude}. Figure~\ref{fig:delay} shows some results. The CPT in Fig.~\ref{fig:delay}(a) leads to the double attosecond pulse in Fig.~\ref{fig:delay}(b) and features a slow modulation corresponding to the pulse separation of 97\,as. This separation can be continuously tuned. Subfigures (c) and (d) show the same for a pulse separation of 194\,as. Additionally, the relative phase of the subpulses can be adjusted, shown by subfigures (e) and (f).	

In summary, we have shown that quasi-phase matching HHG with a shaped counterpropagating pulse train enables control over the spectral amplitude and phase of the harmonics, including elimination of the attosecond chirp. The concept may be applied to any implementation of partial phase matching that permits longitudinal addressing of the modulation frequency, including grating-assisted phase matching \cite{Cohen-2007-Grating} which has the potential for high efficiency extension to keV photon energies.

%

\end{document}